\documentclass[twocolumn,aps,prc,superscriptaddress,showpacs,floatfix]{revtex4}
\usepackage{amssymb}
\usepackage{amsmath,bm}
\usepackage[normalem]{ulem}
\usepackage[dvips]{color}
\usepackage{booktabs}
\usepackage{threeparttable}
\usepackage{graphicx}
\usepackage{CJK}
\renewcommand{\sout}{\bgroup \color[rgb]{1,0,0}\ULdepth=-.5ex \ULset}

\begin{document}

\title{The symmetry energy in nucleon and quark matter}

\author{Lie-Wen Chen\footnote{email: lwchen$@$sjtu.edu.cn}}
\affiliation{School of Physics and Astronomy and Shanghai Key Laboratory for
Particle Physics and Cosmology, Shanghai Jiao Tong University, Shanghai 200240, China}
\affiliation{Center of Theoretical Nuclear Physics, National Laboratory of Heavy Ion
Accelerator, Lanzhou 730000, China}
\date{\today}

\begin{abstract}
The symmetry energy characterizes the isospin dependent part of the equation of state of isospin
asymmetric strong interaction matter and it plays a critical role in many issues of nuclear physics and astrophysics.
In this talk, we briefly review the current status on the determination of the symmetry
energy in nucleon (nuclear) and quark matter. For nuclear matter, while the subsaturation density
behaviors of the symmetry energy are relatively well-determined and significant progress
has been made on the symmetry energy around saturation density, the determination of the
suprasaturation density behaviors of the symmetry energy remains a big challenge.
For quark matter, which is expected to appear in dense matter at high baryon densities,
we briefly review the recent work about the effects of quark matter symmetry energy on the
properties of quark stars and the implication of possible existence of heavy quark stars on
quark matter symmetry energy. The results indicate that the $u$ and $d$ quarks could feel very different
interactions in isospin asymmetric quark matter, which may have important implications on
the isospin effects of partonic dynamics in relativistic heavy-ion collisions.
\end{abstract}
\maketitle

\section{Introduction}  

The density dependence of nuclear matter symmetry energy $E_{\text{sym}}(\rho )$, which
characterizes the isospin dependent part of the equation of state (EOS) of asymmetric
nuclear matter, attracts much interest in current research frontiers of nuclear physics
and astrophysics.
The exact information on nuclear matter symmetry energy is critically
important for understanding many challenging questions ranging from
the structure of radioactive nuclei, the reaction dynamics induced by rare isotopes,
the liquid-gas phase transition in asymmetric nuclear matter and the isospin dependence
of QCD phase diagram to the location of neutron drip line and r-process paths in the nuclear
landscape, the properties of neutron stars and the explosion mechanism of supernova as well as
the frequency and strain amplitude of gravitational waves from inspiraling neutron star binaries~\cite{LKB98,Lat04,Ste05,Bar05,CKLY07,LCK08,Tra12,Tsa12,Lat12a,LiBA12,LiBA14,Hor14,ChenLW14,WangR15,Bal16,LiZX14,JiangWZ14,DongJM14,ZhangYX16,XuC16}.
In addition, some interesting issues of
possible new physics beyond the standard model~\cite{Hor01b,Sil05,Wen09,ZhengH14} may also be related
to the symmetry energy.

While important progress has been made in recent years on constraining the
density dependence of nuclear matter symmetry energy,
mainly based on the data analysis from terrestrial laboratory measurements and astrophysical
observations~\cite{Ste05,Bar05,CKLY07,LCK08,Tra12,Tsa12,Lat12a,LiBA12,LiBA14,Hor14,ChenLW14,Bal16,LiZX14,JiangWZ14,DongJM14,ZhangYX16,XuC16}
as well as some {\it ab initio} theoretical calculations~\cite{Car15},
large uncertainties still exist, especially for its high
density behaviors~\cite{Xia09,Fen10,Rus11,XuC10b,WangYJ15}.
Accurate determination of
the density dependence of nuclear matter symmetry energy thus
provides a strong motivation for the investigation of isospin nuclear physics
at the new/planning radioactive isotope beam facilities around the world, such
as CSR/Lanzhou and BRIF-II/Beijing in China, SPIRAL2/GANIL in France, FAIR/GSI
in Germany, RIBF/RIKEN in Japan, SPES/LNL in Italy, RAON in Korea, and
FRIB/NSCL and T-REX/TAMU in USA.

At extremely high baryon densities, the matter would become deconfined quark
matter. Since isospin symmetry is still satisfied in quark matter, quark
matter symmetry energy is thus involved for the properties of isospin
asymmetric quark matter.
The isospin asymmetric quark matter could exist in compact stars such as neutron
stars or quark stars, and it could be also produced in
heavy ion collisions induced by neutron-rich nuclei at ultra-relativistic energies.
Although essential
progress has been made in understanding the density
dependence of nuclear matter symmetry energy, little empirical information is
known on the density dependence of quark matter symmetry energy~\cite{DiToro06,DiToro10,Pag10,Sha12,ChuPC14,LiuH16,Xia16}.
Theoretically, it is hard to evaluate the quark matter symmetry energy since the
{\it ab initio} Lattice QCD simulations do not work at large baryon densities
while model-independent perturbative QCD works only at extremely
high baryon densities.

In the present paper, we briefly review the current status on the
determination of the symmetry energy in nuclear and quark matter.

\section{The symmetry energy}

\label{EOS}

The EOS of isospin asymmetric nuclear matter with baryon density
$\rho =\rho _{n}+\rho _{p}$ ($\rho _{n}$ and $\rho _{p}$ denote
the neutron and proton densities, respectively)
and isospin asymmetry $\delta =(\rho _{n}-\rho _{p})/(\rho _{p}+\rho _{n})$,
given by the binding energy per nucleon, can be expanded in $\delta $ as
\begin{equation}
E(\rho ,\delta )=E_{0}(\rho )+E_{\mathrm{sym}}(\rho )\delta ^{2}+O(\delta
^{4}),  \label{EOSANM}
\end{equation}%
where $E_{0}(\rho )=E(\rho ,\delta =0)$ represents the EOS of
symmetric nuclear matter, and the nuclear matter symmetry energy is expressed as
\begin{equation}
E_{\mathrm{sym}}(\rho )=\frac{1}{2!}\frac{\partial ^{2}E(\rho ,\delta )}{%
\partial \delta ^{2}}|_{\delta =0}.  \label{Esym}
\end{equation}%
In Eq.~(\ref{EOSANM}), the disappearance of odd-order terms in $\delta $ is due to
the exchange symmetry between protons and neutrons in nuclear matter when
the electromagnetic interaction among nucleons is not considered.
Neglecting the contribution from higher-order $\delta $ terms in Eq.~(\ref{EOSANM})
leads to the well-known empirical parabolic law for the EOS of
asymmetric nuclear matter, which has been verified by many many-body theories
to date, at least for densities up to moderate values \cite{LCK08,Cai12}.

Around a reference density $\rho _{r}$, the $E_{\mathrm{sym}}(\rho )$ can
be expanded in $\chi_r=(\rho -{\rho _{r}})/\rho _{r}$ as
\begin{equation}
E_{\text{sym}}(\rho )=E_{\text{sym}}({\rho _{r}})+\frac{L(\rho _{r})}{3}\chi_r+O(\chi_r^2),
\end{equation}
where $L(\rho _{r})=3{\rho _{r}}\frac{\partial E_{\mathrm{sym}}(\rho )}{\partial
\rho }|_{\rho ={\rho _{r}}}$ is the density slope parameter at $\rho _{r}$ which characterizes
the density dependence of the symmetry energy around $\rho _{r}$.
At saturation density $\rho_0$ of symmetric nuclear matter,
$L(\rho _{r})$ is reduced to $L \equiv L(\rho _{0})$.

Similarly as for nuclear matter, the EOS of three-flavor $u$-$d$-$s$
quark matter with baryon number density $n_B$, isospin asymmetry $\delta_q $
and $s$-quark number density $n_s$, defined by the binding energy per baryon number,
can be also expanded in isospin asymmetry $\delta_q $ as
\begin{equation}
E(n_B ,\delta_q, n_s)=E_{0}(n_B, n_s)+E_{\mathrm{sym}}(n_B, n_s)\delta_q ^{2}+\mathcal{O}(\delta_q ^{4}),
\label{EOSAQM}
\end{equation}%
where $E_{0}(n_B, n_s)=E(n_B ,\delta_q =0, n_s)$ is the binding energy per
baryon number in isospin-symmetric $u$-$d$-$s$ quark matter with equal
$u$ and $d$ fraction; the quark matter symmetry energy
$E_{\mathrm{sym}}(n_B, n_s)$ is expressed as
\begin{equation}
E_{\mathrm{sym}}(n_B, n_s) =\left. \frac{1}{2!}\frac{\partial ^{2}E(n_B
,\delta_q, n_s)}{\partial \delta_q ^{2}}\right\vert _{\delta_q =0}.
\label{QMEsym}
\end{equation}%
The isospin asymmetry of quark matter is defined as
\begin{equation}
\delta_q = 3\frac{n_d-n_u}{n_d+n_u},
\label{delta}
\end{equation}
which equals to $-n_3/n_B$ with the isospin density $n_3 = n_u-n_d$ and
$n_B = (n_u+n_d)/3$ for two-flavor $u$-$d$ quark matter. We note that the
above definition of $\delta_q $ for quark matter has been extensively used
in the literature~\cite{DiToro06,Pag10,DiToro10,Sha12,ChuPC14,LiuH16,Xia16}, and one has
$\delta_q = 1$ ($-1$) for quark matter converted by pure neutron (proton) matter
according to the nucleon constituent quark structure, consistent with the
conventional definition for nuclear matter, namely, $\frac{\rho_n -\rho_p}{\rho_n +\rho_p}=-n_3/n_B$.
In Eq.~(\ref{EOSAQM}), the absence of odd-order terms in $\delta_q $ is due to
the exchange symmetry between $u$ and $d$ quarks in quark matter when
the electromagnetic interaction among quarks is not considered. The higher-order
coefficients in $\delta_q $ are shown to be very small in various model
calculations~\cite{ChuPC14} and thus the empirical parabolic law is also approximately
satisfied for the EOS of asymmetric quark matter.

\section{Nuclear matter symmetry energy around the saturation density}

\begin{figure}[tbp]
\includegraphics[scale=0.31]{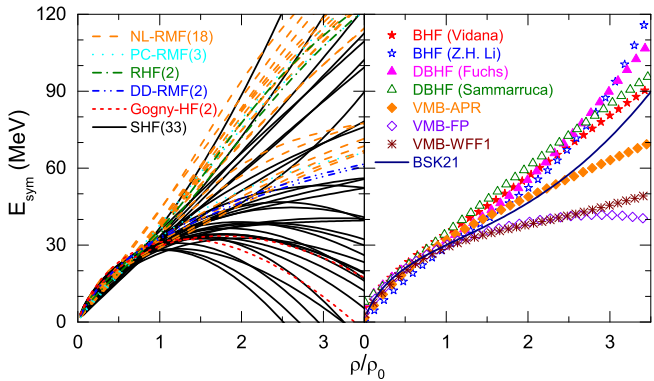}
\caption{(Color online) Nuclear matter symmetry energy as a function of the density
normalized by the corresponding saturation $\rho_0$ from various energy
density functionals with $60$ interactions (Left panel) and microscopic
many-body approaches (Right panel). See text for the details.}
\label{Fig1EsymSys}
\end{figure}

Theoretically, it remains a big challenge to calculate
the density dependence of nuclear matter symmetry energy, mainly due to
our poor knowledge of nuclear (effective) interactions as well as the
limitation of the present nuclear many-body techniques.
As an example,
shown in the left panel of Fig.~\ref{Fig1EsymSys} is the nuclear matter symmetry energy as a function
of the density normalized by the corresponding saturation density
$\rho_0$ with $60$ well-calibrated interactions in various energy density
functionals, namely,
$18$ nonlinear RMF interactions (FSUGold, PK1s24, NL3s25, G2, TM1, NL-SV2,
NL-SH, NL-RA1, PK1, NL3, NL3*, G1, NL2, NL1, IU-FSU, BSP, IUFSU*, TM1*),
$3$ point-coupling RMF interactions (DD-PC1, PC-PK1, PC-F1),
$2$ relativistic HF interactions (PKO3 and PKA1),
$2$ density-dependent RMF interactions (DD-ME1 and DD-ME2),
$2$ Gogny interactions (D1S and D1N), and
$33$ Skyrme interactions (v090, MSk7, BSk8, SKP, SKT6,
SKX, BSk17, SGII, SKM*, SLy4, SLy5, MSkA, MSL0, SIV, SkSM*, kMP, SKa, Rsigma,
Gsigma, SKT4, SV, SkI2, SkI5, BSK18, BSK19, BSK20, BSK21, MSL1, SAMi, SV-min,
UNEDF0, UNEDF1, TOV-min).
These interactions include the $46$ interactions used in Ref.~\cite{Roc11}
(except BCP which is designed for density up to only $0.24$ fm$^{-3}$)
and other $14$ interactions (i.e., BSK18, BSK19, BSK20, BSK21, MSL1, SAMi, SV-min,
UNEDF0, UNEDF1, TOV-min, IU-FSU, BSP, IUFSU*, TM1*) constructed more recently.
For more details, see Ref.~\cite{ChenLW15}.
Shown in the right panel of Fig.~\ref{Fig1EsymSys} are the results from some microscopic many-body
approaches, namely, the non-relativistic Brueckner-Hartree-Fock (BHF) approach~\cite{Vid09,Li08}, the
relativistic Dirac-Brueckner-Hartree-Fock (DBHF) approach~\cite{Kla06,Sam10}, and
the variational many-body (VMB) approach~\cite{APR98,Fri81,Wir88}. The result from
the Skyrme interaction BSk21 is also included in the right panel of Fig.~\ref{Fig1EsymSys} for comparison.
It is clearly seen that
various theoretical models and approaches predict very different density behaviors
of the symmetry energy, especially at supra-saturation densities, indicating
the importance of experimental constraints on the density dependence of
nuclear matter symmetry energy.

During the last decade, a number of experimental or observable probes have
been proposed to constrain the density dependence of nuclear matter
symmetry energy~\cite{LCK08}.
Most of them are for the symmetry energy around saturation density
while a few of probes are for the supra-saturation density behaviors.
By analyzing the data from nuclear reactions, nuclear structures,
and the properties of neutron stars based on various models or methods,
more than $30$ constraints, mainly on the parameters
$E_{\text{\textrm{sym}}}(\rho _{0})$ and $L$, have been obtained,
see, e.g. Refs.~\cite{LiBA12,ChenLW14} for a summary. As indicated in
Ref.~\cite{ChenLW14}, these constraints on
$E_{\text{\textrm{sym}}}(\rho _{0})$ and $L$ cannot be equivalently
reliable since some of them do not have any overlap at all.
However, essentially all the constraints seem to agree with
$E_{\text{\textrm{sym}}}(\rho _{0}) = 32.5 \pm 2.5$ MeV and
$L = 55 \pm 25$ MeV.\\

\begin{figure}[tbp]
\includegraphics[scale=0.4]{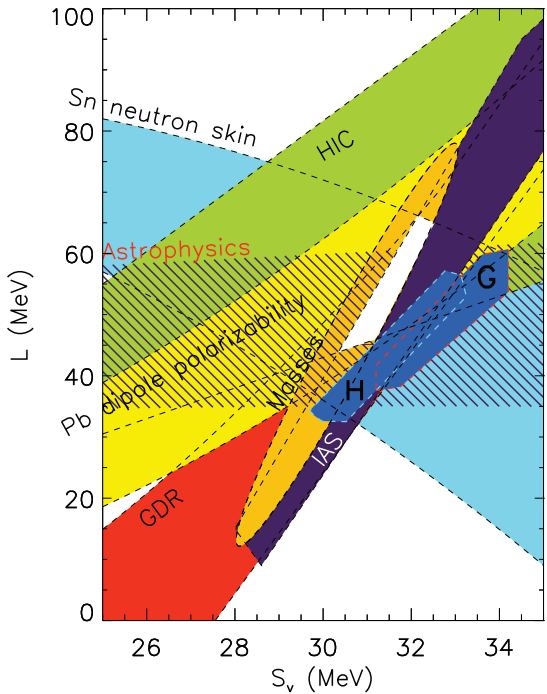}
\caption{(Color online) Constraints on symmetry energy slope
parameter $L$ and its magnitude $S_v$ at $\rho_0$ from six experimental
analyses and one astrophysical analysis.
See the text for further discussion.
G and H refer to the results from the neutron matter studies of
Hebeler et al.~\cite{Heb10} and Gandolfi et al.~\cite{Gan12}, respectively.
Taken from Ref.~\cite{Lat14}.}
\label{Fig2EsymSat}
\end{figure}

The experimental probes of nuclear matter symmetry energy,
especially those related to the structure properties
of finite nuclei, generally depend on both
$E_{\text{\textrm{sym}}}(\rho _{0})$ and $L$. This is because
the structure properties of finite nuclei are generally related to
the subsaturation density behaviors of the symmetry energy due to
the surface existence of finite nuclei, and a constraint of the
symmetry energy at subsaturation density will generally lead to
a correlation between $E_{\text{\textrm{sym}}}(\rho _{0})$ and $L$
at saturation density~\cite{ChenLW11,ZhangZ13}.
Therefore, the constraints on nuclear matter symmetry energy
around saturation density are usually plotted in the
$E_{\text{\textrm{sym}}}(\rho _{0})$-$L$ plane.
As an example, shown in Fig.~\ref{Fig2EsymSat} are six typical experimental constraints
on $E_{\text{\textrm{sym}}}(\rho _{0})$ (denoted as $S_v$ in Fig.~\ref{Fig2EsymSat})
and $L$ summarized by Lattimer and Steiner~\cite{Lat14}, namely,
the constraint
deduced from nuclear masses taken from Hartree-Fock calculations with
the UNEDF0 density functional~\cite{Kor10} (labeled ``nuclear mass''),
the constraint
from Skyrme-Hartree-Fock analysis on the neutron skin thickness
of Sn isotopes by Chen et al.~\cite{ChenLW10} (labeled ``Sn neutron skin''),
the constraint
from the electric dipole polarizability $\alpha_D$ of $^{208}$Pb measured
by Tamii et al.~\cite{Tam11} (labeled ``dipole polarizability''),
the constraint
from the centroid energy of the giant
dipole resonance for $^{208}$Pb taken from Trippa, Colo and
Vigezzi~\cite{Tri08} (labeled ``GDR''),
the constraint
from an improved quantum molecular dynamics (ImQMD) transport model
analysis~\cite{Tsa09} of the isospin diffusion data from two different
observables and the ratios of neutron and proton
spectra in collisions at $E/A=50$ MeV involving $^{112}$Sn
and $^{124}$Sn (labeled ``HIC''), and
the constraint
from the energies of excitations to isobaric analog
states taken from Danielewicz and Lee~\cite{Dan14} (labeled ``IAS'').
For more details on these constraints, see, Ref.~\cite{Lat14}.
The white region displayed in Fig.~\ref{Fig2EsymSat} represents
the consensus agreement of the six experimental
constraints above, giving quite precise values
of $E_{\text{\textrm{sym}}}(\rho _{0}) = 31.5 \pm 1.0$ MeV
and $L = 55 \pm 11$ MeV.

It is interesting to see from Fig.~\ref{Fig2EsymSat} that while other experimental constraints
put a positive correlation between $E_{\text{\textrm{sym}}}(\rho _{0})$
and $L$, the constraint from Sn neutron skin gives a negative
correlation.
This negative correlation is particularly intresting since combing it
with other constraints will significantly improve the constraint
on $E_{\text{\textrm{sym}}}(\rho _{0})$ and $L$ simultaneously.
As pointed out by Zhang and Chen~\cite{ZhangZ13}, the negative
correlation can be understood from the fact that
the neutron skin thickness of heavy
nuclei is uniquely fixed by the symmetry energy density slope $L(\rho)$ at a
subsaturation cross density $\rho_c \approx 0.11$ fm$^{-3}$ ($\approx 2/3\rho_0$) rather than at saturation
density, and a fixed value of $L({\rho_c })$ can generally lead to a
negative $E_{\text{sym}}({\rho _{0}})$-$L({\rho _{0}})$ correlation~\cite{ZhangZ13}.

Also included in Fig.~\ref{Fig2EsymSat} is
a constraint
from astrophysical observation based on Bayesian analysis on
the currently available neutron star mass and radius measurements
by Steiner and Gandolfi~\cite{Ste12} (labeled ``astrophysics''),
for which a phenomenological parametrization for EOS of neutron
matter near and above the saturation density is used
with partial parameters determined by quantum
Monte Carlo calculations.
In addition, Fig.~\ref{Fig2EsymSat} also includes the results deduced from neutron
matter constraints by Hebeler et al. (labeled ``H'')~\cite{Heb10} and by
Gandolfi et al.~\cite{Gan12} (labeled ``G''). It is seen that the regions
deduced from neutron matter deviate from the white region in
Fg.~2 and this could be due to the fact that the higher-order
contributions have been neglected in deducing the
symmetry energy from pure neutron matter, as pointed out by
Lattimer and Steiner~\cite{Lat14}. It should be noted that
although the empirical parabolic law works well
for the evaluation of the symmetry energy magnitude, it might
not be a good approximation for the calculation of the density slope
of the symmetry energy for which the derivative with respective
to the density is involved.

\section{Nuclear matter symmetry energy at subsaturation densities}

In recent years, significant progress has been made in determining nuclear
matter symmetry energy at subsaturation densities. In particular,
it has been well established that the data of binding energy of finite nuclei
can put rather stringent
constraints on nuclear matter symmetry energy at a subsaturation density
$\rho \approx 2/3\rho_0$~\cite{Hor01a,Fur02,Wan13,Dan14,ZhangZ13,Bro13}.
It should be noted that a fixed value of $E_{\text{sym}}({\rho})$ at
subsaturation density constrained from nuclear binding energy (nuclear mass)
can generally lead to a positive $E_{\text{sym}}({\rho _{0}})$-$L({\rho _{0}})$
correlation~\cite{ChenLW11,ZhangZ13}, as observed in Fig.~\ref{Fig2EsymSat}.
For the symmetry energy at even lower densities around $\rho \approx \rho_0/3$,
it has been shown recently~\cite{ZhangZ15} that the data of the
electric dipole polarizability $\alpha_D$ in $^{208}$Pb can put a quite accurate
constraint. This can be easily understood from the following interesting
relation~\cite{ZhangZ15}
\begin{equation}
\label{AlphaDAsym}
\alpha_{\mathrm{D}}(A) = \frac{e^2}{24} \frac{ A \left\langle r^2 \right\rangle}{a_{\mathrm{sym}}(\frac{27}{125}A)},
\end{equation}
which suggests that the $\alpha _ {\text{D}}$ of a nucleus with mass
number $A$ is inversely proportional to the symmetry energy coefficient $a_{\mathrm{sym}}$
of a nucleus with mass number $(\frac{3}{5})^3A$. For $^{208}$Pb, one has
$\alpha_{\mathrm{D}}(A=208)\varpropto 1/a_{\mathrm{sym}}(A=45)$.
From the strong correlation between the $a_{\mathrm{sym}}(A)$
and the $E_{\mathrm{sym}}(\rho_A )$ at a specific density
$\rho_{A}$~\cite{Cen09,ChenLW11,Dan14,Ala14}, the $\alpha_{\mathrm{D}}$ in
$^{208}$Pb is thus expected to be strongly correlated with $E_{\mathrm{sym}}(\rho )$ at
$\rho = \rho_{A=45} \approx \rho_0/3$~\cite{Ala14}.

\begin{figure}[tbp]
\includegraphics[scale=0.36]{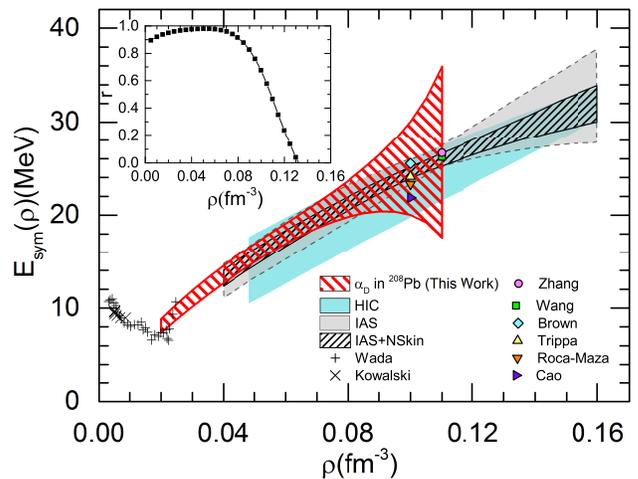}
\caption{(Color online) Constraints on the symmetry energy magnitude $E_{\mathrm{sym}}(\rho)$ as
a function of density $\rho$ (see text for the details). The inset shows the density
dependence of the Pearson correlation coefficient $r$ between $1/\alpha_{\mathrm{D}}$
in $^{208}$Pb and the $E_{\mathrm{sym}}(\rho)$. Taken from Ref.~\cite{ZhangZ15}.}
\label{Fig3EsymSubSat}
\end{figure}

The red hatched band shown in Fig.~\ref{Fig3EsymSubSat} represents the constraints
on the symmetry energy obtained from analyzing the data $\alpha_D$ in $^{208}$Pb.
Shown in the inset in Fig.~\ref{Fig3EsymSubSat} is the density dependence of the Pearson
correlation coefficient $r$ between the symmetry energy and the $\alpha_D$ in $^{208}$Pb.
At very low densities (i.e., less than about $0.02$ fm$^{-3}$) where the fraction of light
clusters becomes significant, clustering effects significantly increase
the symmetry energy~\cite{Typ10}, and thus the constraints on the symmetry energy at
densities below $0.02$~fm$^{-3}$ are not shown in Fig.~\ref{Fig3EsymSubSat} although the $r$ value is
still large. At higher densities (e.g., above $0.11$ fm$^{-3}$), the $r$ value becomes much
smaller and effective constraints cannot be obtained.\\

Also shown in Fig.~\ref{Fig3EsymSubSat} are the constraints from transport model analyses of
mid-peripheral heavy ion collisions of Sn isotopes (HIC)~\cite{Tsa09} and the SHF analyses
of isobaric analogue states (IAS) as well as combing additionally the neutron skin
``data'' (IAS+NSkin) in Ref.~\cite{Dan14}, and six constraints on the value of $E_{\text{sym}}(\rho)$
around $2/3\rho_0$ from binding energy difference between heavy isotope pairs
(Zhang)~\cite{ZhangZ13}, Fermi-energy difference in finite nuclei (Wang)~\cite{Wan13}, properties
of doubly magic nuclei (Brown)~\cite{Bro13}, the giant dipole resonance in $^{208}$Pb (Trippa)~\cite{Tri08},
the giant quadrupole resonance in $^{208}$Pb (Roca-Maza)~\cite{Roc13a} and the soft dipole excitation
in $^{132}$Sn (Cao)~\cite{Cao08}.
These constraints consistently favor a relatively
soft symmetry energy or EOS of asymmetric nuclear matter at
subsaturation densities.

Furthermore, Fig.~\ref{Fig3EsymSubSat} displays the experimental results of
the symmetry energies at very low densities (below $0.2\rho_0$) with temperatures in the
range $3\sim11$ MeV from the analysis of cluster formation in heavy ion collisions
(Wada and Kowalski)~\cite{Nat10}.
It is interesting
to see that the constrained $E_{\text{sym}}(\rho)$ around $\rho_0/7$ from $\alpha_D$ in $^{208}$Pb
is nicely consistent with the results extracted from heavy ion collisions (Wada)
which consider the clustering effects~\cite{Nat10}, and this feature seems
to indicate that the clustering effects do not affect much the symmetry
energy above about $\rho_0/7$ since the mean-field calculations of
$\alpha_D$ in $^{208}$Pb do not include the clustering effects.
In addition, we note the $E_{\text{sym}}(\rho)$ has been predicted
in microscopic calculations (see, e.g., Refs.~\cite{APR98,Dri14,Wel15}) and the results are
consistent with the experimental constraints shown in Fig.~\ref{Fig3EsymSubSat}.

\section{Nuclear matter symmetry energy at supra-saturation densities}

While the subsaturation density behaviors of nuclear matter symmetry energy have been
relatively well-determined and considerable progress has been made on
constraining the symmetry energy around the saturation density, the supra-saturation
density behavior of the symmetry energy remains elusive and largely
controversial. FOPI data on the
$\pi^- /\pi^+$ ratio in heavy-ion collisions favor
a quite soft symmetry energy at $\rho \ge 2\rho_0$ from the isospin and momentum
dependent IBUU04 model analysis~\cite{Xia09} while an opposite conclusion has
been favored from the improved isospin dependent quantum molecular dynamics
(ImIQMD) model analysis~\cite{Fen10}. A recent analysis of FOPI $\pi^- /\pi^+$
ratio data based on the isospin dependent Boltzmann-Langevin
approach~\cite{Xie13} supported the conclusion of the IBUU04 model analysis.
A careful check is definitely needed to understand the above model dependent conclusion.
Recently, Xu \textit{et al.}~\cite{XuJ13} explored the pion in-medium
effect on the $\pi^- /\pi^+$ ratio in heavy-ion collisions at various energies
within the framework of a thermal model, and they demonstrated that
the pion in-medium effects reduce the $\pi^- /\pi^+$ ratio in
heavy-ion collisions compared to that using free pions, especially at lower
incident energies. Therefore, to understand quantitatively the symmetry energy
effect on $\pi^- /\pi^+$ ratio in heavy-ion collisions, it is important to include
the isospin-dependent pion in-medium effects~\cite{Hon14,Son15}, although this is highly nontrivial
in the implementation of the transport model simulations.

\begin{figure}[tbp]
\includegraphics[scale=0.45]{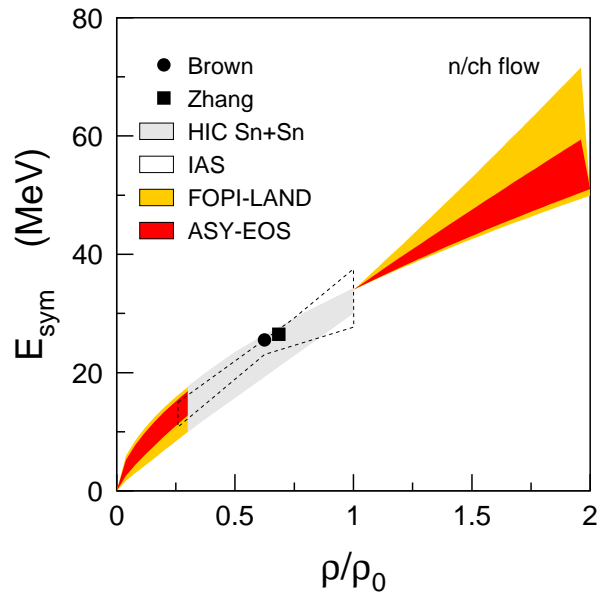}
\caption{(Color online) Constraints on the density dependence of the symmetry energy
deduced from the FOPI-LAND result of Ref.~\protect\cite{Rus11} and the ASY-EOS
result of Ref.~\protect\cite{Rus16}.
The lower density results of Refs.~\protect\cite{Tsa09,Bro13,ZhangZ13,Dan14} are
given by the symbols, the grey area (HIC), and the dashed contour (IAS).
For clarity, the FOPI-LAND and ASY-EOS results are
not displayed in the interval $0.3 < \rho/\rho_0 < 1.0$.
Taken from Ref.~\cite{Rus16}.}
\label{Fig4EsymSupraSat}
\end{figure}

By analyzing the elliptic flow ratio of neutrons to protons or light
nuclei from the FOPI/LAND data for $^{197}$Au + $^{197}$Au
collisions at 400 MeV/nucleon within the UrQMD model, Russotto
\textit{et al}.~\cite{Rus11} obtained a moderately soft symmetry energy with a
density dependence of the potential part proportional to
$(\rho/\rho_0)^{\gamma}$ with $\gamma = 0.9 \pm 0.4$, which is shown in
Fig.~\ref{Fig4EsymSupraSat} as a yellow band. In a more recent work,
Cozma \textit{et al}.~\cite{Coz13} analyzed the FOPI/LAND data of
neutron-proton elliptic flow difference and ratio for $^{197}$Au + $^{197}$Au
collisions at 400 MeV/nucleon within the Tubingen QMD model by using a
parametrization of the symmetry energy derived from the momentum dependent Gogny
force, and they extracted a moderately stiff symmetry energy.
Most recently, Russotto
\textit{et al}.~\cite{Rus16} obtained a new and more stringent constraint on the
symmetry energy for the regime of supra-saturation density with a considerably smaller
uncertainty, i.e., $(\rho/\rho_0)^{\gamma}$ with $\gamma = 0.72 \pm 0.19$ (see the red
band in Fig.~\ref{Fig4EsymSupraSat}), based on the UrQMD
model analysis on the FOPI/LAND data of the elliptic flow ratio of neutrons
to charged particles for $^{197}$Au + $^{197}$Au
collisions at 400 MeV/nucleon. As shown in Fig.~\ref{Fig4EsymSupraSat}, this new constraint of a softer
symmetry energy in the regime of supra-saturation density seems to be also
consistent with the obtained constraints at subsaturation densities
as shown in Fig.~\ref{Fig3EsymSubSat}.\\

Besides using heavy ion collisions to constrain the supra-saturation density
behavior of the symmetry energy, it has been also proposed recently~\cite{ChenLW15}
that the three bulk characteristic parameters $E_{\text{\textrm{sym}}}({\rho _{0}})$,
$L$ and the density curvature parameter $K_{\mathrm{sym}} =9\rho _{0}^{2}\frac{d^{2}E_{\mathrm{sym}}(\rho )}{%
\partial \rho ^{2}}|_{\rho =\rho _{0}}$ essentially determine the symmetry energy with
the density up to about $2\rho_0$. This opens a new window to constrain the
supra-saturation density behavior of the symmetry energy from its precise knowledge
around saturation density.

\section{Quark matter symmetry energy}

The investigation of quark matter symmetry energy has been just started
in recent years and there are essentially no any empirical information
on the density dependence of quark matter symmetry energy.
Within the confined-isospin-density-dependent-mass
(CIDDM) model~\cite{ChuPC14}, it has been shown recently that
the isovector properties of quark matter may play an important role in
determining the properties of strange quark matter
and quark stars. In particular, if the recently discovered heavy
pulsars PSR J1614-2230~\cite{Dem10} and PSR J0348+0432~\cite{Ant13}
with mass around $2M_{\odot}$ were quark stars, they can put
important constraint on the quark matter symmetry energy.\\

\begin{figure}[tbp]
\includegraphics[scale=0.75]{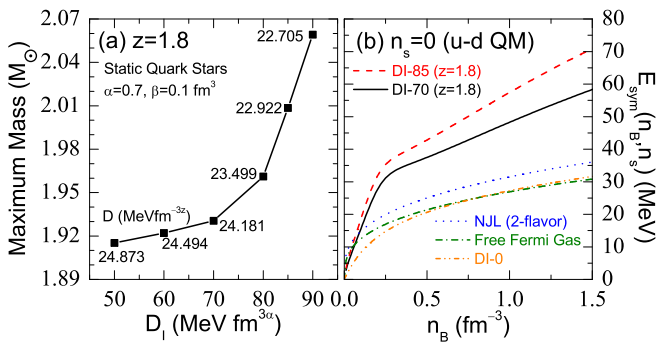}
\caption{(Color online) Left panel: $D_I$ dependence of the maximum mass of static quark stars
in the CIDDM model with $z=1.8$.
Right panel:
The symmetry energy of two-flavor $u$-$d$ quark matter as a function of baryon
number density in the CIDDM model with DI-70 ($z=1.8$) and DI-85 ($z=1.8$).
For comparison, the results of DI-0 as well as the symmetry energy of a free
quark gas and normal quark matter within
conventional NJL model are also included.}
\label{Fig5QMEsymMRz}
\end{figure}

In the CIDDM model, the equivalent
quark mass in isospin asymmetric quark matter with baryon density
$n_B$ and isospin asymmetry $\delta $ is parameterized as~\cite{ChuPC14}
\begin{eqnarray}
m_q &=& m_{q_0} + m_{I} + m_{iso} \notag \\
&=& m_{q_0} + \frac{D}{{n_B}^z} - \tau_q \delta {D_I}n_B^{\alpha}e^{-\beta n_B},
\label{mqiso}
\end{eqnarray}
where $m_{q0}$ is the quark current mass and $m_I = \frac{D}{{n_B}^z} $ reflects
the isospin-independent part of the quark interactions in quark matter,
$z$ is the quark mass scaling parameter, $D$ is a parameter
determined by stability arguments of SQM; $D_I$, $\alpha $, and $\beta $ are parameters describing isospin
dependence of the quark-quark effective interactions in quark matter, $\tau_q $ is the
isospin quantum number of quarks and here we set $\tau_q = 1$ for $q=u$ ($u$ quarks),
$\tau_q = -1$ for $q=d$ ($d$ quarks), and $\tau_q = 0$ for $q=s$ ($s$ quarks).
Shown in the left panel of Fig.~\ref{Fig5QMEsymMRz} is the
$D_I$ dependence of the maximum mass of static quark stars.
The value of the $D$ parameter at different $D_I$ shown in Fig.~\ref{Fig5QMEsymMRz}
corresponds to the value at which the quark star maximum mass
becomes largest. It is seen from the left panel of Fig.~\ref{Fig5QMEsymMRz}
that the maximum mass of quark stars is sensitive to the $D_I$
parameter and it increases with $D_I$.
To obtain a quark star with mass larger than $1.93M_{\odot }$, we find
the minimum value of $D_I$ should be $70$ MeV$\cdot$fm$^{3\alpha}$, and the
corresponding parameter set is denoted as DI-70 ($z=1.8$). For DI-70 ($z=1.8$),
we have $D_I = 70$ MeV$\cdot$fm$^{3\alpha}$, $\alpha=0.7$, $\beta=0.1$ fm$^3$,
$D=24.181$ MeV$\cdot$fm$^{-3z}$, and $z=1.8$.

Shown in the right panel of Fig.~\ref{Fig5QMEsymMRz} is the density dependence of
two-flavor $u$-$d$ quark matter symmetry energy in the CIDDM model with DI-70 ($z=1.8$).
For comparison, we also include the results of the DI-85 ($z=1.8$) parameter set (which produces
a quark star mass of $2.01M_{\odot }$), the DI-0 parameter set (i.e., $D_I=0$)
as well as the symmetry energy of a free quark gas (with $m_{u0} = m_{d0} = 5.5$ MeV)
and normal quark matter within conventional Nambu-Jona-Lasinio
(NJL) model~\citep{Reh96}. One can see that the DI-0 and NJL model predict
a very similar quark matter symmetry energy as that of the
free quark gas, while the DI-70 ($z=1.8$) parameter set predicts a two times larger quark
matter symmetry energy than the free quark gas. Our results thus indicate that
the two-flavor $u$-$d$ quark matter symmetry energy should be at least about twice that
of a free quark gas or normal quark matter within conventional NJL model in order to
describe the PSR J1614-2230 and PSR J0348+0432 as quark stars.
It should be mentioned that the symmetry
energy of two-flavor color superconductivity (2SC) phase is
about three times that of the normal quark matter phase~\citep{Pag10}, and thus is
close to the symmetry energy predicted by DI-70 ($z=1.8$).

\section{Summary}

We have given a brief overview of the current status on the symmetry energy
in nuclear and quark matter.
For nuclear matter, considerable progress has been made in determining the
density dependence of nuclear matter symmetry energy during the last decade
due to the great efforts from experiments, astrophysical observations and
theoretical calculations in the community.
Very encouragingly, one can see that
the various constraints on the symmetry energy at subsaturation
densities are consistently convergent and the the symmetry energy values
at subsaturation densities have been constrained with good precision.
Around saturation density,
although the values of the symmetry energy magnitude
$E_{\text{\textrm{sym}}}({\rho _{0}})$
and its density slope $L$ at
saturation density can vary largely depending on the data or models, all the constraints
obtained so far from nuclear reactions, nuclear structures, and the properties
of neutron stars are consistent with $E_{\text{\textrm{sym}}}({\rho _{0}}) = 32.5 \pm 2.5$ MeV
and $L = 55 \pm 25$ MeV.
On the other hand, all the constraints on nuclear matter symmetry energy at
supra-saturation densities have been obtained from analyzing data in heavy-ion collisions within
transport models, and these constraints are largely controversial.
More high quality data and more accurate theoretical
methods are definitely needed to further reduce the theoretical and
experimental uncertainties of the constraints on the density dependence
of nuclear matter symmetry energy.

The quark matter symmetry energy is a new topic and it may play an important role
in understanding properties of isospin asymmetric quark matter, which could be formed
or exist in ultra-relativistic heavy-ion collisions induced by neutron-rich nuclei and
in the interior of neutron stars or quark stars. Based on a quark matter model, i.e., the
confined-isospin-density-dependent-mass model, it has been shown that if the recently
discovered pulsars with two times solar mass were quark stars, the two-flavor $u$-$d$
quark matter symmetry energy should be at least about twice that of a free quark gas
or normal quark matter within conventional NJL model. This result indicates that the
$u$ and $d$ quarks could feel very different interactions in isospin asymmetric quark matter,
which may have important implications on the isospin effects of partonic dynamics in
heavy-ion collisions at ultra-relativistic energies, e.g., at RHIC in USA, FAIR in Germany,
and NICA in Russia.

\section*{Acknowledgments}
The author would like to thank Bao-Jun Cai,
Rong Chen, Peng-Cheng Chu, Wei-Zhou Jiang, Che Ming Ko, Bao-An Li, Kai-Jia
Sun, Rui Wang, Xin Wang, De-Hua Wen, Zhi-Gang Xiao, Chang Xu, Jun Xu,
Gao-Chan Yong, Zhen Zhang, and Hao Zheng for fruitful collaboration and
stimulating discussions. This work was supported in part by the Major State Basic Research
Development Program (973 Program) in China under Contract Nos.
2013CB834405 and 2015CB856904, the National Natural Science
Foundation of China under Grant Nos. 11625521, 11275125 and
11135011, the Program for Professor of Special Appointment (Eastern
Scholar) at Shanghai Institutions of Higher Learning, Key Laboratory
for Particle Physics, Astrophysics and Cosmology, Ministry of
Education, China, and the Science and Technology Commission of
Shanghai Municipality (11DZ2260700).\\


\end{document}